# What can neuronal populations tell us about cognition?

Iñigo Arandia-Romero[1], Ramon Nogueira[1], Gabriela Mochol[1], and Rubén Moreno-Bote[1,2]

[1] Center for Brain and Cognition & Department of Information and Communications Technologies, University Pompeu Fabra, 08018 Barcelona, Spain
[2] Serra Húnter Fellow Programme, 08018 Barcelona, Spain

Corresponding author:
Rubén Moreno-Bote
Email: ruben.moreno@upf.edu

Contributions: All authors discussed and wrote the paper

Acknowledgements: This work has been supported by the Spanish PSI2013-44811-P and SlowDyn FLAGERA-PCIN-2015-162-C02-02 grants from MINECO to R. M. B., and IJCI-2014-21937 grant from MINECO to G. M.

# Highlights

- Neuronal population (NP) level analysis is slowly pervading systems neuroscience
- NPs provide both unprecedented temporal and spatial resolution to study decision-making
- However, large NPs come with challenging data analysis and interpretative problems
- Challenges can be alleviated by using models of NP activity

# Abstract

Nowadays, it is possible to record the activity of hundreds of cells at the same time in behaving animals. However, these data are often treated and analyzed as if they consisted of many independently recorded neurons. How can neuronal populations be uniquely used to learn about cognition? We describe recent work that shows that populations of simultaneously recorded neurons are fundamental to understand the basis of decision-making, including processes such as ongoing deliberations and decision confidence, which generally fall outside the reach of single-cell analysis. Thus, neuronal population data allow addressing novel questions, but they also come with so far unsolved challenges.

## Introduction

Single-neuron electrophysiology has provided golden ages in neuroscience passing from the discoveries of Lord Adrian [1] to those by Hubel and Wiesel [2]. This approach relies on finding mappings between the activity of a single neuron and external world, body or internal variables, most often summarized with the so-called 'tuning curve', that is, the mean neuron response as a function of the relevant variable. As such, single-cell electrophysiology has proven to be a central pillar over which many theories of the brain rest [2-8].

In the last 20 years, there has been a surge of new technologies that allow recording the activity of hundreds of cells within and across brain areas, either with dense electrodes or with imaging techniques [9-16] and pioneering work has shown the benefits of these neuronal population (NP) recordings [17-22]. New emerging technologies even promise several orders of magnitude improvement in the number of recorded cells, leaping from hundreds to thousands of cells [23, 24]. However, these promises come with new problems. The most prominent challenges arise in experimental settings that involve animal behavior, where the number of trials is limited to avoid animal satiation and exhaustion, and to minimize overtraining to better study naturalistic behaviors. Considering the activity of each neuron as one dimension in the neuronal state space, how should we analyze and model multi-dimensional neuronal activity with just a few tens of trials per condition? How can we benefit from the many dimensions that NP offers? And, most importantly, what are the new questions that can be addressed with NPs about cognitive processes that cannot, or at least hardly, be asked with single neurons?

We review recent literature that can help to answer the above questions with the focus on decision-making. We provide examples of a better understanding of the transformation from stimulus to choices by using the activity of simultaneously recorded cells. We argue about the difficulty, and even impossibility, of drawing qualitatively similar conclusions using single-cell electrophysiology. Then, we talk about the challenges that we face with large NPs and limited number of trials, as those expected from experiments that combine both animal behavior and NPs recordings. Finally, we point to past and more recent work that could benefit from using NP recordings and discuss possible promising future directions.

## Advantages of NPs

The success of NP approaches has been clearly exemplified in brain-machine interface applications [25-28]. Recent studies have shown how it is possible to restore hand control in a quadriplegic patient [29] or aid non-human primates in walking [30] by reading out the population activity of motor cortex and converting these signals into motor commands bypassing injured nerves. 'Reading out' or 'decoding' means computing an online estimate of the value of an external world or internal variable from the activity of the recorded neurons [31-33]. For instance, decoding the population activity of V1 neurons with linear classifiers can predict with high accuracy the orientation of a stimulus drifting grating in a trial-by-trial basis [20, 34, 35]. Recent literature tends to favor approaches based on decoding over information theory [19, 20, 34, 36-43], mostly because decoding techniques can be applied with relatively small number of trials while information-theoretic methods require computing probability distributions over as many dimensions as neurons in the NP, a problem daunted by the curse of dimensionality [33]. In addition, decoding techniques provide trial-by-trial estimates of external and internal variables that can be directly interpreted and compared to their actual values [44].

Despite the success of NP approaches to understand the basics of attention [36], working memory [45] and decision-making [37, 38, 46], they are sometimes contemplated with suspicion. Indeed, in some applications it could be argued that NPs do not offer radically new knowledge compared to single-neuron approaches. For instance, while decoding analysis can provide better estimates of information content in a NP by taking into account the trial-by-trial correlated noise among cells, rough estimates can be obtained by extrapolating the information of singles neurons to small NPs of a few tens of cells [47] (but not for populations larger than 100-1000 cells due to the presence of 'differential correlations' [48]). Moreover, research based on NP recordings sometimes ends up using traditional single-neuron-based analysis [49, 50]. What are the unique opportunities that NPs can offer? Here we give examples of how NPs can provide new knowledge that is very hard - and in some cases impossible - to acquire from single-neuron-based analysis.

There are at least two broad families of problems that cannot be addressed with single-neuron approaches: 1) internal neuronal dynamics that is not time-locked to observable stimulus or body variables, and 2) variables that are encoded at the population level and not at the single-neuron level. The first category relates to neuronal processing that is internally or externally triggered without experimental control. These phenomena might include trial-by-trial fluctuations of attention that are not cued experimentally, mind wandering, and any other neuronal processing that is not time-locked to the stimulus or body variables [34, 36, 37, 39, 41, 45, 51-63]. Because of their lack of significant correlation to observable experimental variables, these phenomena cannot be fully studied using single-neuron approaches. The second category relates to neuronal processing that requires coordination of neurons in a NP. For instance, stable representation of a world variable, such as world-centered body location of mice [64], can occur despite unstable representation of that variable at the single-cell level (Fig. 1a). Similarly, theoretical work shows that stable representation of working memory signals can be achieved in NPs despite unstable single-neuron representations [65]. Thus, observing single neurons can lead to the erroneous conclusion that the encoding of the variable is unreliable, while the reality is different, namely, the variable is reliably encoded at the NP level. Further, theoretical studies suggest that the encoding of variables can involve temporal delays between neurons without concomitant changes of activity in any single cell [66] (Fig. 1b), and conclusions on information content drawn from combining single cells recordings can be misleading if there is shared response variability among cells in the NP [32, 48] (Fig. 1c).

## Opportunities from NPs to understand decision-making

NPs can provide new ways to study the processes underlying decision-making and cognition. These novel opportunities with NP arise because crucial decision-making variables might not be generally representable by single neurons. . An example of such a variable is 'decision confidence' [67-71], that is, the confidence of a decision maker about her choice being correct based on accumulated noisy evidence [68]. A large family of mathematical models for two-alternative decision-making tasks called 'race models' requires at least two types of neurons racing to generate the choice [67, 68]. Here each racing neuron represents at each time point the evidence for each option accumulated so far (Fig. 2). For this family of models, decision confidence is encoded in the difference between the activity of two racing neurons (in addition to elapsed time), the so-called 'balance of evidence', a quantity that cannot be represented by the activity of neither of the neurons alone [68, 72]. The activity of any single racing neuron has little information about confidence. For example, it is possible to have no change in firing rate with confidence (Fig. 2a,b, cell 1) or even increase in firing rate (Fig.2, cell 2) for lower confidence (lower panels of Fig 2a vs Fig. 2b). More realistic models based on probabilistic population codes [73], which incorporate the Poisson-like nature of neuronal responses, also represent both choices and confidence simultaneously in the difference in activity of different neuronal pools [74].

Recent experimental work has shed new light into how changes of mind might be encoded in neuronal populations during decision-making, a process that can also guide further studies in the encoding of decision confidence on NPs. Kiani and colleagues recorded the activity of large NPs in area 8Ar of prearcuate gyrus while macaque monkeys performed a motion discrimination task [37]. The authors used decoding techniques to find a hyperplane in multidimensional activity space that separated well animals' choices on a trial-by-trial basis (Fig. 3a). Importantly, the distance of the population activity vector from the hyperplane could be used to infer in which trials the animal 'changed its mind' (Fig 3b). This method allows studying the ongoing process of choice at unprecedented levels of temporal resolution. Decision confidence theories predict that confidence should be related to the same NP activity distance, but these predictions remain to be tested.

A recent paper by Rich and Wallis has studied the deliberation process in economic-based choices at a surprising level of temporal resolution using up to 16 electrodes in areas 11 and 13 in the monkey orbitofrontal cortex [38]. Following a smart experimental design, a linear decoder learned to discriminate between different offers presented on forced-choice single-offer trials, showing how offers were encoded in the NP when they were presented separately. This decoder was then used without retraining to predict the internal deliberation process in free-choice trials where two offers were simultaneously available, uncovering alternations over time between the states representing the available options (Fig. 3c). The untested assumption in this work is that the representation of offers is the same in forced and free choice trials. However, consistently with this assumption, longer total duration of the state corresponding to the chosen offer was observed compared to the unchosen one. Thus, the activity of NPs reveals a rich dynamic representation of choice options, which would be very hard to obtain with single neurons. Similarly complex dynamics has been also reported in other tasks and areas during decision-making [41, 45].

## Challenges from NPs

The above examples illustrate the importance of NP analysis. But what is the NP size that we ought to record from to address a given question? A priori, there is no limit to the desired size: the larger, the better, although in some cases with just a bunch of simultaneously recorded cells it is possible to address questions that are unfeasible with just a single cell, as described above for relevant theories of decision confidence. As a rule of thumb, the larger the NP, the more precisely the state of the neuronal network can be assessed [34, 53, 75], the more accurately stimulus or body variables can be decoded [20, 34, 35], and the more variables can be studied simultaneously [29, 30].

However, the promises of population-based analysis do not come free of challenges. The most important challenge is model specification given large number of parameters and limited number of trials. Consider, for instance, a population of 100 neurons with 200 trials per distinct condition and a binary classification task (400 trials in total). Using a quadratic discriminant for this task requires estimating a number of parameters that scales as the square of the number of neurons, around 5 thousand parameters, which is outside the reach of this limited amount of data. A linear classifier could be still used in this condition because 400 trials is in general sufficient to estimate its 101 free parameters. However, consider now a population of 1000 neurons with again 200 trials per condition and the same binary classification task. In this case, even for a linear classifier, we need to estimate 1001 parameters, which is above the number of data points collected. Roughly speaking and based on our experience, we would need at least around 5 times more trials per class and per neuron recorded to train a linear classifier (~5×neurons×classes), which for large NPs will certainly be difficult to achieve. The problem is even bigger for more complex decoders with many more free parameters, such as deep feedforward networks [76]. Deep learning approaches that have boosted machine learning in the last few years can be hardly suitable for neuronal data based on a few trials per neuron and condition [76]. Novel approaches that would work with standard 'few-trials' data need to be developed.

What approaches are then suitable for 'few-trials' data? This is a domain ripe for research, and here we discuss two recently used methods. The first one is to use linear decoders together with regularization techniques [37, 52, 77]. In this way, parameters that provide none or little explanatory power are pushed to zero or very small values, therefore reducing the complexity of the decoder and thus dampening overfitting. Regularization has been shown to be useful for NPs approximately as large as 150 neurons with more than 500 trials per condition [37]. However, a possible danger is that regularization implicitly introduces priors, which constraint the family of solutions, but might turn to be wrong. The second method consists of building models with increasing complexity starting from basic principles and performing model comparison using cross-validation techniques [34, 78, 79]. Basic principles refer here to the knowledge that has been acquired across many years of research. For instance, knowledge about translational invariance of real-world objects incorporated into convolutional networks allows to significantly reduce their number of free parameters, thus improving dramatically their classification performance [76]. Another example of prior knowledge that can be included in decoders is the Poisson-like nature of neuronal responses [4, 73, 80-83] and the gain modulations induced by global fluctuations in population activity [34, 51, 54, 75, 78, 84]. For instance, a recent model of NP activity with Poisson-like firing and both heterogeneous multiplicative and additive gains has been shown to explain better macaque V1 responses than alternative models ('multi-gain' model [34]) (Fig. 4). The first important advance is that this analysis directly exploits NP data, as otherwise it is impossible to study the modulation of single-neuron activity with global gain. Secondly, the resulting model has culminated research that has shown that global gain modulations can be either purely additive [51, 75], purely multiplicative [84], homogeneous multiplicative and additive [54], or a mixture of homogeneous multiplicative and heterogeneous additive gains [78] (Fig. 4).

Finally, it is important to avoid some potential pitfalls when using NPs. First, as larger NPs are available, computing 'noise correlation' matrices (correlations between all pairs of neurons at fixed stimulus conditions, [32, 85, 86]) will become unfeasible, because the covariance matrix becomes singular when the number of neurons exceeds the number of trials [87], unless strong priors are used [88]. There is also a tendency to build large surrogate NPs from independently recorded single-neurons [63, 89, 90]. Surrogate NPs can be useful in some special circumstances [63, 89-91], but overusing them could lead to erroneous conclusions. This can be the case when neuronal populations contain 'differential correlations', which can be very small and yet have a huge impact on information and computations, and are absent in surrogate populations [48].

## Conclusions: what can NPs offer in the future?

We have shown that NP analysis is slowly pervading neuroscience, but fulminating strides are expected to occur soon. First, ongoing debates on the role of certain brain areas in decision-making and on its neuronal mechanisms will dramatically benefit from measuring the activity of very large NPs [92-95]. Second, decision-making is characterized by complex behavior that in real-life conditions involve a myriad of variables, such as complex stimuli and body variables. The many-fold dimensions of neuronal activity will allow studying the simultaneous encoding of those variables in the brain. Dimensionality reduction techniques [96], including decoders, will leave room to new methods (e.g., [97]) more suitable when the complexity of the experimental design does not permit for such reductionist, low dimensional, view of neuronal activity, as computational neural network models predict [98-100]. Finally, it is also expected that new-generation decoders and other approaches suitable for 'few-trials' datasets, typical for behavioral and systems neuroscience, will allow precise estimates of sensory information from increasingly large NPs [48, 86].

# Figures and legends

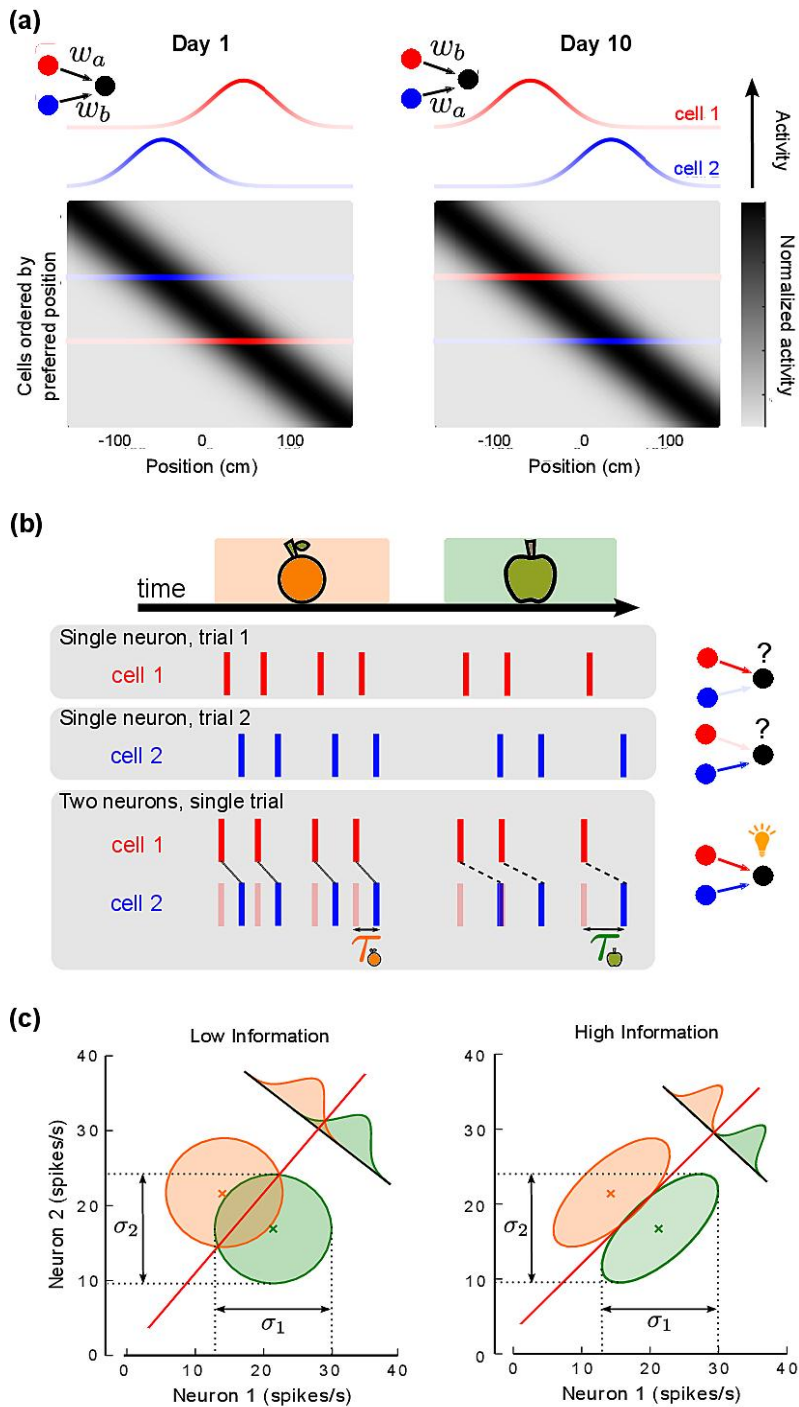

**Figure 1**: Neuronal populations (NPs) are necessary to understand the neuronal code.
 (**a**) NPs can represent stable information while the tuning of individual neurons changes [64]. The activity of cell 1 (red) and cell 2 (blue) for different positions can change from the first day (top-left) to the tenth day (top-right), while the information represented by the NP remains constant (bottom). The ordering of the cells by preferred position differs from the first (bottom-left) to the tenth (bottom-right) day due to changes in individual tuning, but jointly they represent the same map. To recover the information represented by the NP a readout neuron (black circle at the top) adapts the weights ($w_i$-s) to the tuning of each neuron (red and blue circles) at a given day. The information recovered across days is the same.

(**b**) Stimulus information in theory can be encoded in the relative spike timing between neurons [66]. Vertical bars represent the spiking activity of cell 1 (red) and cell 2 (blue) in three cases while two stimuli (orange and apple) are presented consecutively (top). The first two cases show the activity recorded from single cells. Here, the readout neuron (black circle on the right) is not able to recognize the stimulus that has been presented. In the third case, in contrast, the readout neuron can measure the difference in the timing of spikes from cell 1 and cell 2, and recover the stimulus by using the relative spike latencies.

(**c**) Pair-wise correlations can help decoding [32]. Each plot shows the mean activity (crosses) and the variability in the response (standard deviation represented by clouds) of two neurons when two different stimuli are presented (orange and green). The responses in the left panel have circular clouds, meaning that the variability is the same in all directions, and therefore the correlation between the two neurons is zero for both stimuli. The panel in the right shows two correlated neurons with the same amount of variability as in the left panel, but spread along the same direction (ellipses). The red line marks the decoding boundary, dividing the space between the area which corresponds to the orange stimulus and to the green one. The clouds overlapping in the left panel mean that in some trials where the green stimulus has been presented the decoder will predict the orange one, and vice-versa. However, for the correlated activity case (right) most of the trials are correctly classified, as the clouds do not overlap. Therefore, predicting the performance of a neuronal code depends on neuronal correlations, a problem that lies beyond single-neuron electrophysiology.

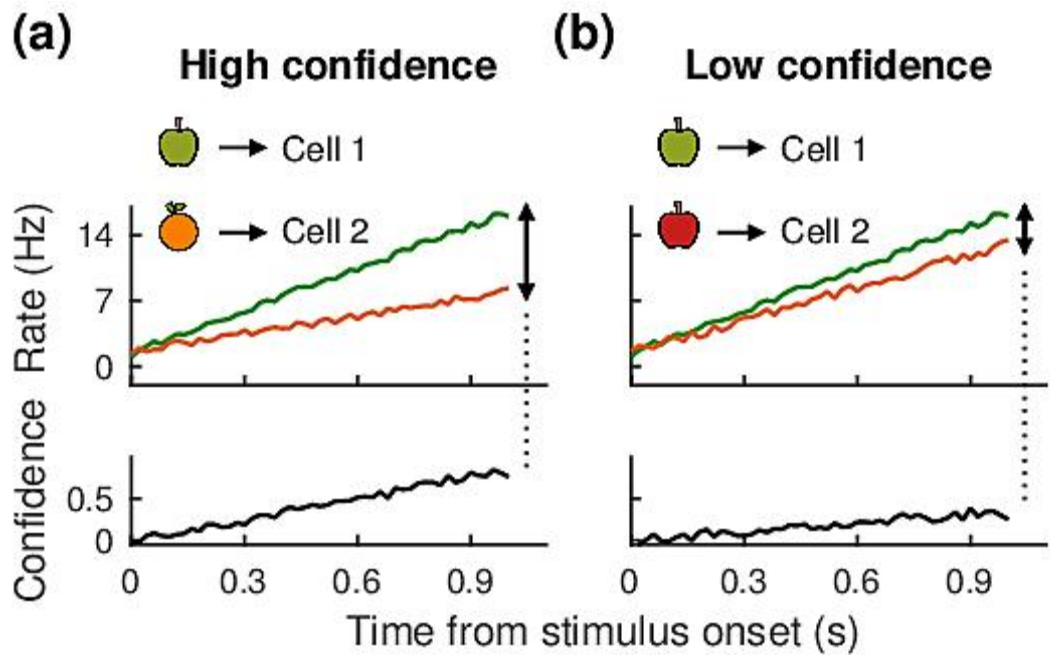

**Figure 2**: Decision confidence in a 'race model'.

(**a**) A decision maker choses between two offers (green apple and orange), whose values are represented by the activity of two neurons (red and green lines), growing over time as more evidence is accumulated. The decision rule is based on what option depicts the highest activity at decision time (dashed vertical line). In this case, the choice was easy because the values of the offers were very distinct, and therefore the confidence on the decision (represented by the distance between the activity of the neurons) was high [68].

(**b**) When the two offers have similar values (red vs green apple), the choice is more difficult, and thus decision confidence (distance between the races) is lower. The estimation of confidence would not be possible if the two neurons were not recorded simultaneously. Comparing the high and low confidence cases (a-b) the activity of the neuron preferring the green apple (cell 1) is identical, and thus this neuron alone does not correlate with decision confidence at the time of the decision. Although the neuron that prefers the lower value offer changes its activity in the two cases, the direction of this change is opposite to the change in confidence (cell 2 has higher rate for the low confidence trial) and cannot give a good approximation of the latter.  nor provide information about the state of the other racing neuron, making confidence estimation impossible from single neuron recordings.

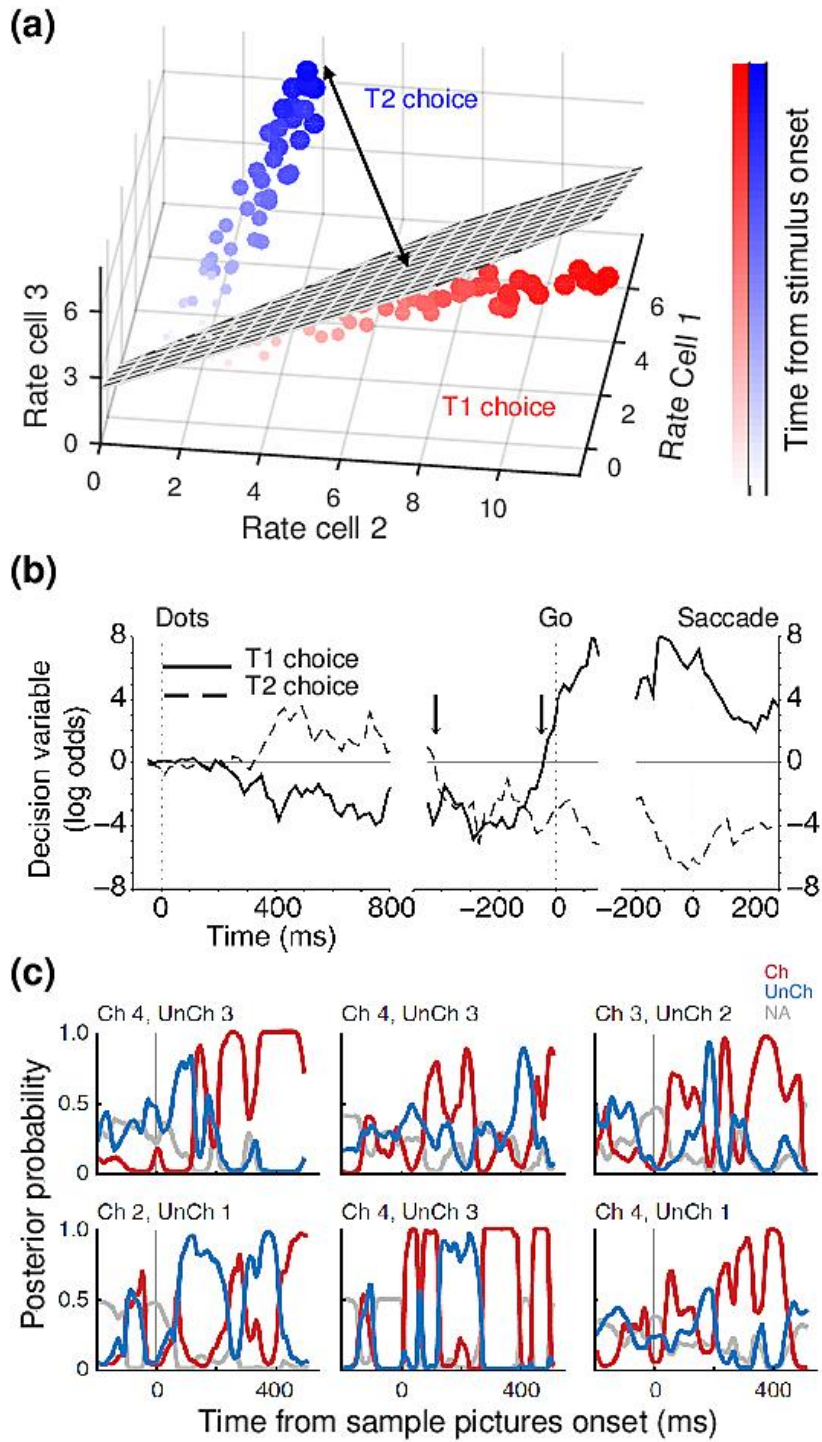

**Figure 3**: Decision making in neuronal response space.
(**a**) The dynamics of decision making of an animal performing a two-alternative task can be visualized in the neuronal response space, that is, the space where each direction corresponds to the activity of a recorded neuron. At a given time window the population activity is a single point in this space (blue and red dots) A linear decoder is trained to find a hyperplane (in this case, a plane in three dimensions) which best separates the points belonging to the two classes (red for T1 choice, and blue for T2 choice). At the beginning of each trial (faded colors) the population activity provides weak evidence about stimulus conditions, and thus the points are close to the hyperplane. As more evidence is accumulated, the population activity diverges away from the hyperplane, making easer to predict the animal's choice. The distance between population activity and the classification hyperplane defines the decision variable (DV, black line).

(**b**) The temporal dynamics of DV reveals 'changes of mind'. In a given trial, the DV crossed the ambiguity point equal to zero (indicated by an arrow), suggesting the presence of a change of mind, that is, an instance in which the hypothetical intended choice flips (two trials are shown). Tracking these changes of mind would not be possible with single-neuron analysis. Reproduced from [37] with the permission of the publisher.

(**c**) Ongoing deliberation inferred NP activity in monkey orbitofrontal cortex. A linear decoder, trained in forced-choice trials, was used to predict choices between simultaneously presented offers in free-choice trials. The posterior probability derived from the decoder for chosen (red), unchosen (blue) and unavailable options (gray, average of both unavailable options), are shown in six different trials. This analysis allows tracking the ongoing deliberation processes at unprecedented temporal resolution. Reproduced from [38] with the permission of the publisher

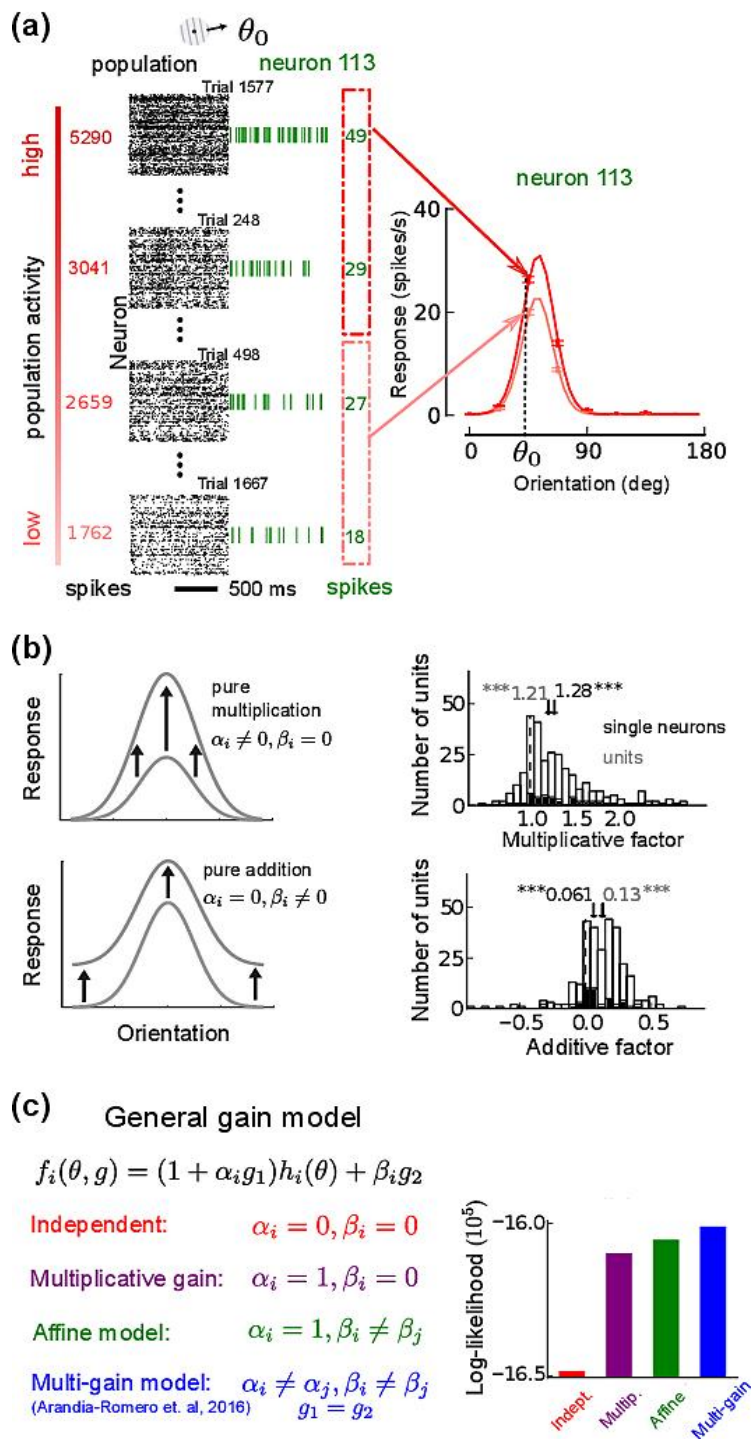

**Figure 4:** Population activity models based on Poisson-like firing and global gain modulation. A large variety of models have been recently used to characterize the activity of NP including the effect of the population activity on single neuronal activity in primary visual cortex (V1) [34, 75, 78, 84].
(**a**) Population activity fluctuates across trials for the same stimulus (left panels). The tuning of a V1 neuron (right) is modulated with population activity (called here global gain), which is estimated as the sum of spikes across all neurons excluding the activity from the neuron for which tuning is being characterized; thus, this analysis cannot be done with single neurons, it requires NPs. Trials were ranked from high (top left) to low (bottom left) global gain for each stimulus condition. The activity of the selected neuron (green spike trains) was averaged across either the top (red box) or bottom (pink box) 50th percentile of trials, and the averages were plotted as a function of stimulus orientation (rightmost panel). The tuning was modulated with global gain (red vs. pink lines). Points and error bars are mean responses and s.e.m., respectively; lines are von Mises fits. Reproduced from [32] with the permission of the publisher.

(b) Global gains can affect the tuning of V1 cells in a multiplicative (top-left) and/or an additive (bottom-left) way. The modulatory effect of the population activity on the tuning curves is heterogeneous, with different multiplicative (top-right) and additive factors (bottom-right) for different neurons. ***: $p < 0.001$. Reproduced from [32] with the permission of the publisher. (**c**) Models of NP activity including global gains. Neurons are assumed to fire as independent Poisson processes (or variations of this, see [78, 84]) conditioned to gain factors $g_1$ and/or $g_2$, which reflect modulatory factors in the NP. The mean firing rate of each cell *i* in the NP is given by the function $f_i(\theta, g)$. The function $h_i(\theta)$ determines the tuning of the cells when the gain factors are zero. The gains can produce multiplicative effects on the tuning curves if $\alpha_i$ is non-zero, and additive effects if $\beta_i$ is non-zero (see panel **b left**)). A recent model for NP activity suggests heterogeneous multiplicative and additive modulations (multi-gain model [34]). The multiplicative gain model assumes that there is not additive gain and that the multiplicative gain is identical for all neurons ($\alpha_i = 1$, $\beta_i = 0$ for all *i*). The affine model allows for heterogeneous additive gains, but the multiplicative gain is still uniform for all neurons ($\alpha_i = 1$ for all *i*). The multi-gain model permits heterogeneous multiplicative and additive gains (it also uses $g_1 = g_2$). The multi-gain model provides better fits of V1 population activity than other models. Cross-validated log-likelihood is shown.